# Spin-Polarized Zener Tunneling in (Ga,Mn)As

E. Johnston-Halperin,[1] D. Lofgreen,[2] R.K. Kawakami,[1] D.K. Young,[1,2] L. Coldren,[2] A.C. Gossard,[2,3] and D.D. Awschalom[1]

[1]*Department of Physics, University of California, Santa Barbara, CA 93106*
[2]*Department of Electrical and Computer Engineering, University of California, Santa Barbara, CA 93106*
[3]*Materials Science Department, University of California, Santa Barbara, CA 93106*

We investigate spin-polarized inter-band tunneling through measurement of a (Ga,Mn)As based Zener tunnel diode. By placing the diode under reverse bias, electron spin polarization is transferred from the valence band of p-type (Ga,Mn)As to the conduction band of an adjacent n-GaAs layer. The resulting current is monitored by injection into a quantum well light emitting diode whose electroluminescence polarization is found to track the magnetization of the (Ga,Mn)As layer as a function of both temperature and magnetic field.

With the recent discovery of ferromagnetic semiconductors compatible with III-V epitaxy[1] the field of spintronics has expanded from all-metal[2] and hybrid metal-semiconductor[3] structures to include all-semiconductor ferromagnetic devices.[4] Such devices have revealed intriguing properties, such as control of their Curie temperature with an applied voltage.[5] This is a consequence of the carrier-mediated nature of the ferromagnetism in these materials, which allows manipulation of the magnetic properties through control of the electronic subsystem (*i.e.* through doping, gating, *etc.*). Here we exploit this control to construct a heavily doped p-n diode, which when operated in the Zener tunneling regime is capable of transferring *electron* spin polarization from the valence band[6] of p-type (Ga,Mn)As to the conduction band of an adjacent n-GaAs layer.

This spin transfer is investigated by monitoring the resulting spin-polarized electron current as it is injected into a (Ga,In)As quantum well (QW) placed in the intrinsic region of an n-i-p light emitting diode (LED). The polarization of the luminescence from the QW is a measure of the spin-polarization of the electron current[7] and is monitored as a function of magnetic field and temperature, revealing that the tunneling process results in a spin-polarization that tracks the magnetization of the (Ga,Mn)As layer. In addition to investigating the tunneling process itself, we are able to circumvent the lack of n-type ferromagnetic semiconductors[1,8] and generate a spin-current in an environment that promises both higher mobility[9] and longer spin-lifetime[10] than is possible in hole mediated transport.

A schematic of the band diagram for such a tunnel diode is presented in Fig. 1a. The device performs as follows: a negative bias is applied between the p-(Ga,Mn)As and the p-GaAs layers, as indicated by the vertical arrow labeled −V. This places the Zener diode tunnel junction (TJ) in reverse-bias and allows electron tunneling from the valence band of the p-(Ga,Mn)As to the conduction band of the adjacent n-GaAs layer. Subsequent to tunneling, the spin polarized electrons are injected into the LED structure (which is under forward bias) and swept into the (Ga,In)As QW where they recombine radiatively with unpolarized holes injected from the p-GaAs layer. As stated above, this will result in a net circular polarization of the luminescence proportional to the spin-polarization of the electron current.[7]

In order to realize this structure, samples are grown in two separate Varian/EPI Gen-II molecular beam eptiaxy (MBE) chambers (chambers A and B). The LED structure is grown first and entirely within chamber A and is comprised of the following layers: 300 nm n-GaAs (Si: $3\times10^{16}$ cm$^{-3}$)/ 30 nm GaAs/ 10 nm $Ga_{0.87}In_{0.13}As$/ 100 nm GaAs/ 1000 nm p-GaAs (C: $5\times10^{18}$ cm$^{-3}$)/ SI-GaAs (100) substrate. The bottom layer (n contact) of the TJ is also grown in chamber A and is comprised of 10 nm n-GaAs (Si: $1\times10^{19}$ cm$^{-3}$). The top layer (p contact) of the TJ device is a p-type $Ga_{0.943}Mn_{0.057}As$ layer,[11] and is grown in chamber B. In order to maintain the quality of the TJ interface during sample transfer, upon completion of growth in chamber A the substrate is cooled to room temperature and an As cap (~2-3 ML) is deposited on the surface. The sample is then removed and subsequently transferred in air to chamber B, where the As cap is desorbed at 400° C in an As atmosphere, and 300 nm of $Ga_{0.943}Mn_{0.057}As$ is grown to complete the TJ structure. The Mn content of the (Ga,Mn)As layer is determined by MnAs RHEED oscillations,[12] the As:Ga beam flux ratio is ~ 25:1, and the growth temperature is stabilized at 260° C using band edge thermometry.[13] Finally, a control sample with 300 nm p-GaAs (C: $1\times10^{20}$ cm$^{-3}$) substituted for the p-(Ga,Mn)As layer is grown entirely within chamber A.

For both optical and electrical measurements the samples are processed into bars ~ 2mm long with cleaved facets, "wedding cake" cross section, and electrical contacts consisting of un-annealed[14] Ti:Au, as shown in Fig. 1b. This structure allows independent electrical access to both the TJ and LED structures, and their respective I-V curves at T = 5 K can be seen in Fig. 1c. A bias of about –0.50 V is required to turn on the TJ, suggesting that the conduction band of the n-GaAs lies close to mid-gap relative to the p-(Ga,Mn)As. The filled squares indicate the I-V characteristic when the drive is applied across the entire sample (as is the case for the following measurements).

The electro-optical measurement technique used here is adapted from previous work in p-i-n heterostructures.[4,15] The samples are placed in the bore of a magneto-optical cryostat capable of reaching fields up to 8 T and temperatures T = 1.2 K - 300 K. The magnetic field is measured *in situ* using a Hall sensor located adjacent to the sample. Luminescence is collected by a collimating lens, passed sequentially through a variable retarder and linear polarizer, dispersed in a 1.33 m spectrometer, and detected with a liquid nitrogen cooled charge coupled device (CCD). Polarization resolution is achieved by measuring the intensity of the QW luminescence while the variable retarder is set to quarter- and three quarter-wave retardance, passing left circularly polarized (LCP) and right circularly polarized (RCP) light respectively. Due to both the effects of quantum confinement and strain on the optical selection rules responsible for the QW luminescence,[16,17] as well as the magnetic anisotropy of the (Ga,Mn)As layer,[1] the measured spin polarization may depend on the collection geometry. We therefore measure both the luminescence collected from the edge of the sample (in plane of QW, easy magnetic axis of (Ga,Mn)As) and through the substrate (out of plane of QW, hard magnetic axis of (Ga,Mn)As).

We first consider the edge-emitted luminescence, i.e. along the +y direction in Fig. 1b. Electroluminescence (EL) is collected from the edge of the sample while holding the drive current constant at −200 µA and T = 5 K. The resulting spectrum can be seen on a semi-log scale in Fig. 2a and shows that the QW luminescence is centered at 1.384 eV and is both sharp (2.70 meV FWHM) and spectrally distinct from the GaAs luminescence (centered at 1.491 eV). We also use SQUID magnetometry to measure the magnetization of the (Ga,Mn)As layer along this direction[18] for an unprocessed sample (Fig. 2b).

In order to determine whether electrical spin injection is taking place, we monitor the polarization of the EL, $P = (I_{LCP} − I_{RCP})/(I_{LCP} + I_{RCP})$, as a function of applied magnetic field. Here $I_{LCP/RCP}$ indicates the intensity of LCP and RCP light respectively integrated over the spectral width of the QW (gray shaded region in Fig. 2a).[19] The results of one such scan taken at T = 5 K and drive current of − 200 µA can be seen in Fig. 2c. We observe no dependence of the polarization on drive current. A background polarization of ~ 3% which does not depend on magnetic field over this range has been subtracted from the data. While the field dependence of the polarization qualitatively mimics that of the magnetization in the (Ga,Mn)As film (Fig. 2b), the coercivity, $H_c$, is found to be somewhat larger (42 G for the EL versus 20 G for the SQUID data). However, as this discrepancy could arise from a variety of extrinsic effects, such as differing shape anisotropy from the as-grown film, increased oxidation due to processing, *etc.*, we do not consider it relevant here.

It is in principle possible that the polarization effects seen in Fig. 2c are an artifact of the luminescence scattering from the dichroic[6] (Ga,Mn)As layer en route to the detector. In order to ascertain the extent of this potential contribution to the field dependence, photoluminescence (PL) is collected with the same measurement geometry as the EL.[4] The linearly polarized excitation beam is incident from the substrate side of the sample (along +z in Fig.1b) and tuned to 1.425 eV (below the GaAs band edge) to minimize optically generated spin polarization. A representative PL spectrum can be seen in the inset to Fig. 2a, and the corresponding field scan is indicated by the open triangles in Fig. 2c. The PL field dependence shows none of the magnetic structure visible in the EL, ruling out both the path dependent effects mentioned above as well as any intrinsic spin splitting in the QW as potential sources for spurious signal. Such spin splitting in the QW could arise from, *e.g.* Zeeman splitting due to fringe fields from the (Ga,Mn)As layer.

The temperature dependence of the remanent polarization (polarization at zero applied field) of the EL may be compared to the temperature dependence of the magnetization. The data can be seen in Fig. 2d, with the EL indicated by filled circles and SQUID data by the solid line. As with the field dependence, the polarization of the EL clearly tracks with the magnetization of the (Ga,Mn)As film.

We also consider emission through the substrate. In this geometry the sample is rotated 90° with respect to the collection optics and applied magnetic field (Fig. 3a). The high saturation field (~ 3 kG) and lack of significant remanence or hysterisis in the magnetization data (Fig. 3b) indicate that this direction is a hard axis for the magnetization of the (Ga,Mn)As film. As with the edge-emission geometry, we monitor the polarization of the EL as a function of applied field at T = 5 K and drive current of – 200 µA. However, due to the relatively large fields required to saturate the magnetization in this geometry, we observe two additional background effects. We measure both a linear variation in the polarization versus magnetic field, and a lifting of the ground state spin degeneracy in the QW resulting in a polarization- and field-dependent Zeeman shift in the energy of the luminescence ($E_{LCP}$ - $E_{RCP}$ ~ 120 µeV at 1 T).

We investigate the spectral dependence of the polarization by reducing the width of the integration window used in calculating the polarization to 1.5 meV and varying this window within the QW luminescence. First, we find that as the integration window moves through the center of the EL, the slope of the linear field dependence mentioned above changes sign. This suggests that the linear background arises from the polarization-dependent Zeeman shifts in the QW energy. Since we observe this effect in both PL studies and measurements of the non-magnetic control sample, we take this behavior to be intrinsic to the QW and therefore subtract it from the data for ease of comparison with the magnetization. Second, we find that while the position of the spectral window does not affect the sign of the spin-injection signal, we do see a modulation in amplitude similar to that seen in the edge geometry.[19]

The resulting field dependence can be seen in Fig. 3c, and shows saturation at 0.81 ± 0.08 % at a field of ~ 3 kG. The amplitude of the saturation is almost identical to that measured in the edge geometry (0.82 ± 0.08 %) and the saturation field is consistent with the magnetization data in Fig. 3b. Initially puzzling, however, is the overall minus sign in the amplitude of the polarization of the substrate-oriented EL. We will address this issue momentarily, but would first like to consider a PL test similar to that described for the edge geometry. Excitation is through the substrate with linearly polarized light at 1.425 eV, and the linear background mentioned above has been subtracted from the field dependence. The data (Fig. 3c, open triangles) again rule out the possibility that the field dependence of the EL originates from either circular dichroism or fringe fields.

In considering the overall minus sign in the substrate-oriented emission data, we examine the selection rules governing recombination in the QW. When binding with a heavy hole (HH) to form an optically active exciton the electron spin lies anti-parallel to the net angular momentum of the exciton, while in the light hole (LH) case it lies parallel. Consequently a given electron spin orientation can give rise to either LCP or RCP light upon recombination, depending on whether it recombines with a light or heavy hole.[20] Therefore, without knowing whether light or heavy holes are involved in the recombination we cannot translate the sign of the luminescence polarization into the electron spin orientation.

However, quantum confinement effects pin the angular momentum of the HH along the growth direction and LH in the plane of the QW.[16,17] As a result, the corresponding excitons must also lie along the growth direction and in the plane, respectively. The sign difference between edge and back emission reported here can then be explained as the recombination of a spin-polarized electron current with light and heavy holes, respectively. Careful analysis of the sign of the measured polarization, coupled with the above assumptions about the exciton specie giving rise to the luminescence, leads us to propose that the direction of this spin-polarized electron current should be parallel to the magnetization of the (Ga,Mn)As film.

As an additional check for potential measurement artifacts, the EL measurements described above were also performed on the non-magnetic control sample. While the background effects are similar to those measured in the magnetic samples, there is no sign of spin injection. To summarize, i) the field, temperature, and orientation dependence of the EL polarization tracks the magnetization of the (Ga,Mn)As layer, ii) PL measurements taken with the same measurement geometry reveal no sensitivity to the (Ga,Mn)As, and iii) a non-magnetic control sample also reveals no significant field dependent polarization. Taken together we find these observations to be compelling evidence that we are measuring a spin-polarized electron current generated by inter-band tunneling from the valence band of the (Ga,Mn)As layer of the TJ.

Measurements of a sample prepared with a 5.6 nm intrinsic GaAs layer inserted in the depletion region of the TJ, effectively increasing the width of the tunnel barrier, reveal only a modest decrease in the remanent polarization ($P_R$ = 0.58%, Fig. 4) and coercivity ($H_c$ = 30 G). The extrinsic effects mentioned above complicate any interpretation of the change in coercivity, but the relative resilience of the remanent polarization is encouraging for the construction of novel band-engineered spintronics devices. We would like to thank J. English and J. Champlain for technical assistance, M.E. Flatté for helpful discussions, and the support of: DARPA/ONR N00014-99-1-1096.

**Figure 1 (a)** Schematic band diagram: solid arrow labeled –V indicates direction of band bending under normal operating conditions. Filled circles represent electrons and open circles holes, arrows indicate spin orientation. Spin polarized holes in the p-(Ga,Mn)As region represent the hole gas thought to mediate ferromagnetism in this material.[1] **(b)** Cartoon representing patterned structure with bias conditions used to obtain the data in (c). Cone in the +y direction indicates luminescence from the QW for edge emission geometry; arrow labeled B indicates direction of the magnetic field. **(c)** I-V curves for the TJ and LED structures as well as for the complete device (TJ + LED) at T = 5 K.

**Figure 2 (a)** EL and PL (inset) spectra at T = 5 K. EL is driven by a current of -200 µA and PL is excited by 1.425 eV linearly polarized light incident through the substrate (PL Pump in Fig. 1b). **(b)** Magnetization data measured in the plane of the sample using a SQUID magnetometer at T = 5 K. **(c)** EL and PL polarization as a function of magnetic field. Excitation conditions are the same as for (a), and gray shaded regions in (a) indicate width of spectral integration. **(d)** Remanent polarization ($P_R$) and remanent magnetization ($M_R$) as a function of temperature. Error bars represent the standard deviation in the individual hysterisis loops.

**Figure 3 (a)** Cartoon of sample orientation for emission through the substrate. Cone indicates QW luminescence and arrow labeled B indicates the direction of the magnetic field. **(b)** Magnetization data measured along the growth direction (hard axis) of the sample at a temperature of 5 K. The slight hysterisis is likely due to a small (< 5°) misalignment of the sample. **(c)** EL and PL as in Fig. 2c.

**Figure 4** EL polarization as a function of field for both a Zener tunneling device similar to the one described above (Zener) and a second sample with a spacer layer inserted into the TJ (Zener + Spacer). Excitation is –200 µA for both samples and T = 5 K.

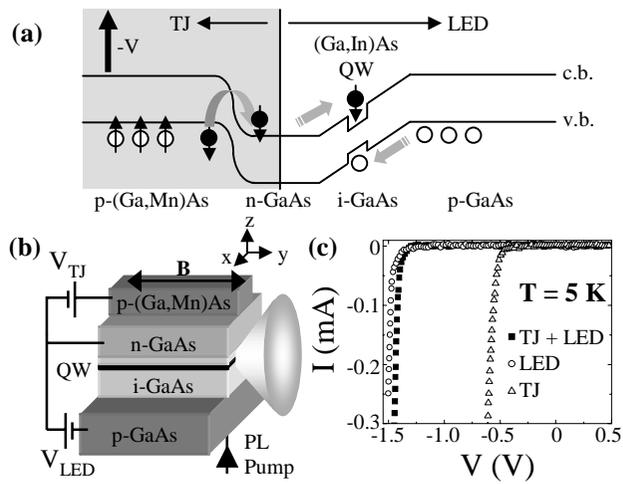

Figure 1

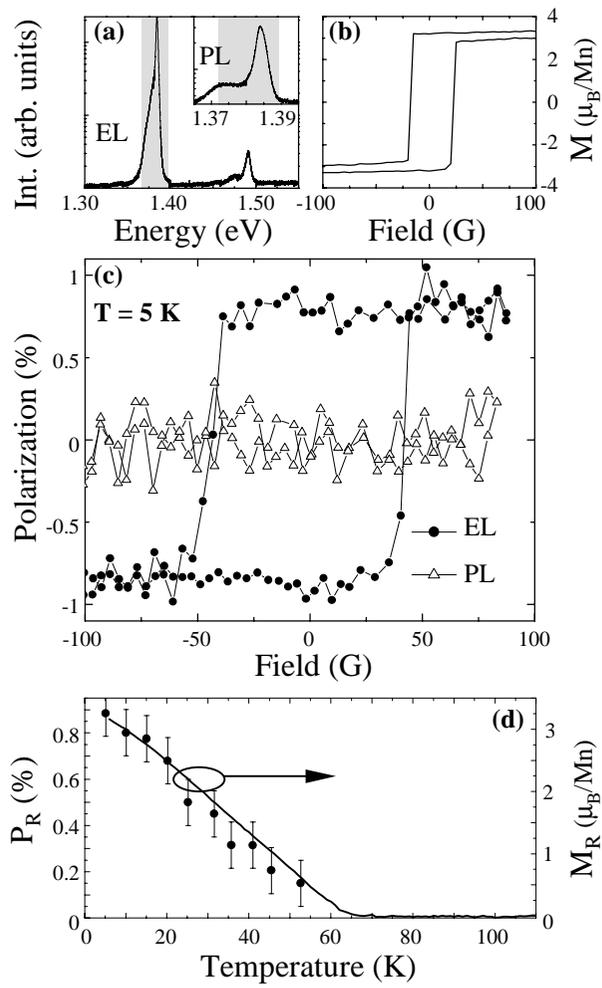

Figure 2

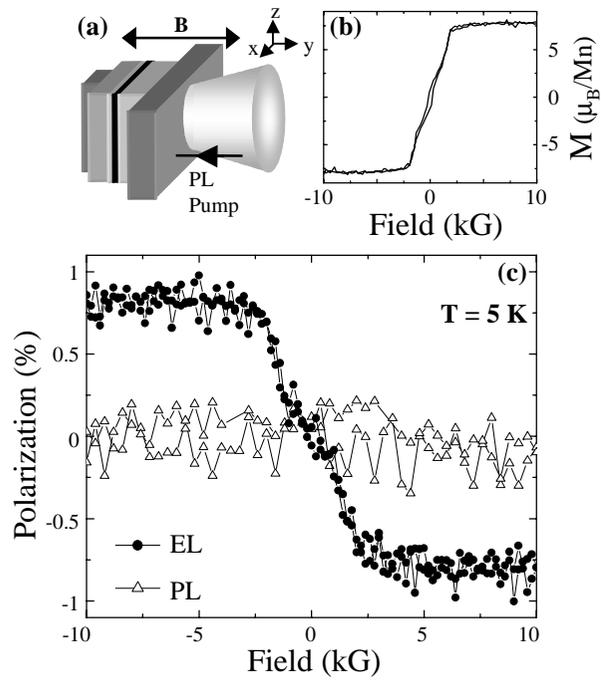

Figure 3

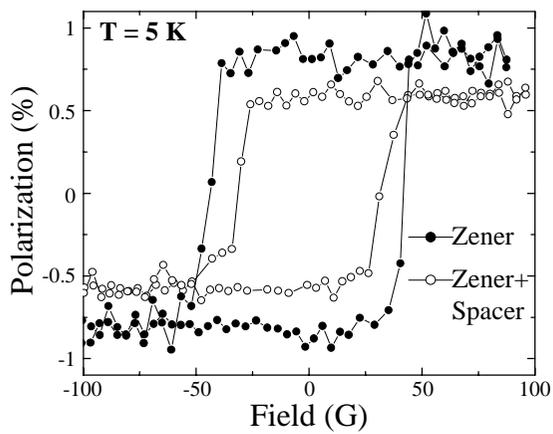

Figure 4